\documentclass[12pt]{article}
\usepackage{latexsym,amsmath,amssymb,amsbsy,graphicx}
\usepackage[cp1251]{inputenc}
\usepackage[T2A]{fontenc}
\usepackage[russian,english]{babel}
\usepackage[space]{cite} % If exist.

\begin{document}
\bigskip

{\bf Magnetic cycles of solar-type stars with different levels of
coronal and chromospheric activity -- comparison with the Sun }

\bigskip

\centerline {E.A. Bruevich, V.V. Bruevich, E.V. Shimanovskaya}

\centerline {\it Sternberg Astronomical Institute, Moscow State
 University,}
\centerline {\it Universitetsky pr., 13, Moscow 119992, Russia}\

\centerline {\it e-mail:  {red-field@yandex.ru, brouev@sai.msu.ru,
eshim@sai.msu.ru} }\

\bigskip

{\bf Abstract.}
%Chromospheric and coronal activity of
%stars of late (F, G, K) spectral classes and also their cyclic
%activity are analyzed based on observational data of solar-type stars
%from the Mount Wilson HK-project. The chromospheric and coronal
%activity of HK-project stars and the Sun is compared to stars
%with active atmospheres observed in Planets Search Programs. It is
%shown that the cyclic activity begins to occur in stars, close by
%their spectral type to the Sun, and becomes more pronounced in
%K-stars. Results of a comparative analysis of
%characteristics of chromospheric and coronal activity of the Sun and
%F, G and K-stars are presented.
The atmospheric activity of the Sun and solar-type stars is analysed
involving observations from HK-project at the Mount Wilson
Observatory, the California and Carnegie Planet Search Program at
the Keck and Lick Observatories, and the Magellan Planet Search
Program at the Las Campanas Observatory. We show that for stars of
F, G and K spectral classes, the cyclic activity, similar to the
11-yr Solar cycles, is different: it becomes more prominent in
K-stars. Comparative study of solar-type stars with different levels
of the chromospheric and coronal activity confirms that the Sun
belongs to stars with the low level of the chromospheric activity
and stands apart among these stars by the minimum level of the
coronal radiation and minimum flux variations of the photospheric
radiation.

\bigskip
{\it Key words:} Sun: activity, stars: solar-type, stars: late-type,
stars: activity, stars: HK-project.

\bigskip

\vskip12pt
\section{Introduction}
{\label{S:intro}}

The study of magnetic activity of the Sun and solar-type stars is of
fundamental importance for astrophysics. This activity of the stars
leads to the complex of composite electromagnetic and hydrodynamic
processes in their atmospheres. Local active regions, which are
characterized by a higher value of intensity of the local magnetic
field, are: plagues  and spots in photospheres, $Ca II$ flocculae in
chromospheres and prominences in coronas.

It is difficult to predict the evolution of each active region in
details. However, it has long been established that the total change
of active areas integrated over the entire solar or stellar disk is
cyclical not only in solar activity but in stellar activity too
(Baliunas et al. 1995; Morgenthaler et al. 2011; Kollath and Olah
2009; Bruevich and Kononovich 2011).

It is well known that the duration of the 11-yr cycle of solar
activity (Schwabe cycle) ranges from 7 to 17 years according to a
century and a half of direct solar observations.

Durations of chromospheric activity cycles, found for 50 stars of
late spectral classes (F, G and K), vary from 7 to 20 years
according to HK-project observations.

The HK-project of Mount Wilson observatory is one of the first and
still the most outstanding program of observations of solar-type
stars (Baliunas et al. 1995; Lockwood et al. 2007). One of the most
important results was the discovery of 11-yr cycles of activity in
solar-type stars.

Currently, there are several databases that include thousands of
stars with measured fluxes in the chromospheric lines of $Ca II$,
see (Wright et al. 2004; Isaacson and Fisher 2010, Arriagada 2011,
Garcia et al. 2010, Garcia et al. 2010), but only for a few tens of
stars the periods of magnetic activity cycles are known (Baliunas et
al. 1995; Radick et al. 1998; Lockwood et al. 2007; Morgenthaler et
al. 2011; Olah et al. 2009).

In our work, we consider the following databases of observations of
solar-type stars with known values of $S$-index:

 1. HK-project -- the program in which the Mount Wilson "S value" was first
defined. The Mount Wilson "S value" ($S$-index, $S$) became the
standard metric of chromospheric activity -- the basic value with
which all future projects of stellar chromospheric activity
observations are compared and calibrated.

2. The California and Carnegie Planet Search Program which includes
observations of approximately 1000 stars at Keck and Lick
observatories  in chromospheric $Ca II$ $H$ and $K$ emission cores.
$S$-indexes of these stars are converted to the Mount Wilson system,
see Wright et al. 2004. From these measurements, median activity
levels, stellar ages, and rotation periods from general
parameterizations have been calculated for 1228 stars, $\sim$ 1000
of which have no previously published S-values.

3. The Magellan Planet Search Program which includes Las Campanas
Observatory chromospheric activity measurements of 670 F, G, K and M
main sequence stars of the Southern Hemisphere. $S$-indexes of these
stars are also converted to the Mount Wilson system, see Arriagada
(2011).

The aims of our paper are: (1) - a study of the place of the Sun
among stars with different levels of chromospheric and coronal
activity belonging to the main sequence on the Hertzsprung-Russell
diagram; (2) - a comparative analysis of chromospheric, coronal and
cyclic activity of the Sun and solar-type stars of F, G and K
spectral classes.

\vskip12pt
\section{The place of the Sun among the solar-type stars with different levels of chromospheric and coronal activity}
{\label{S:place}} \vskip12pt

\begin{figure}[tbh!]
\centerline{
\includegraphics[width=140mm]{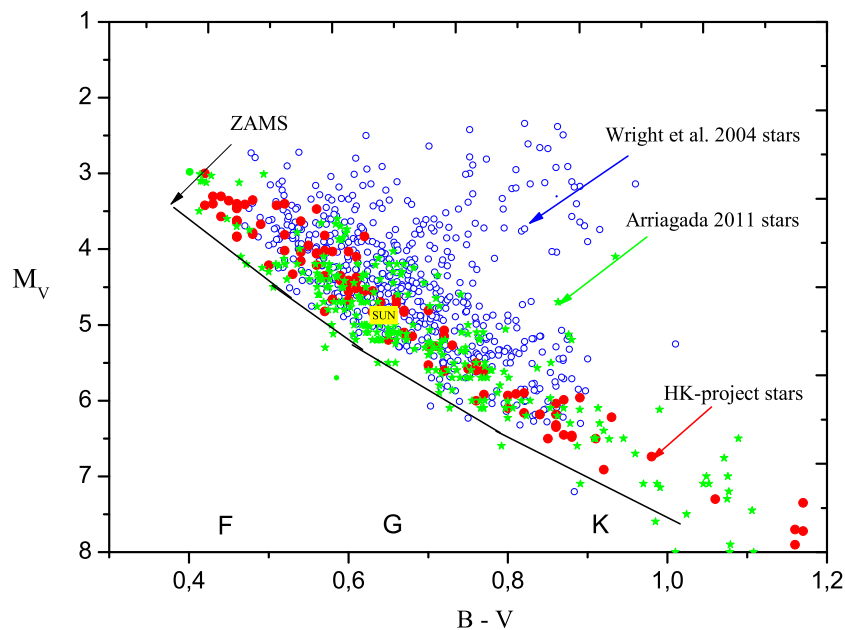}}
 \caption{The Sun among the solar-type stars from different observational
 Programs on the Hertzsprung-Russell diagram.}
{\label{Fi:Fig1}}
\end{figure}

In Figure 1, the Sun and stars from different observed samples are
presented on the Color-Magnitude Hertzsprung-Russell diagram. There
are  1000 solar-type stars from Wright et al. 2004, observed in
Program of Planet Search (blue circles);  660 solar-type stars from
Arriagada (2011) observed in Magellan Planet Search Program (green
circles) and 110 solar-type stars and the Sun from Baliunas et al.
1995, observed in the Mount Wilson HK-project (red circles). The
solid line represents the Zero Age Main Sequence (ZAMS) on the
Hertzsprung-Russell diagram.

Stars which are close to the ZAMS in Figure 1 have the lowest age
among all other stars: log(Age/yr) is about 8 - 8.5. The older the
star is, the farther it is from the ZAMS. Ages of stars in the
Figure 1 varies from $10^8$ to $10^{10}$ years.

We can also see that some stars significantly differ from the Sun by
the absolute magnitude $M_V$ and color indices $(B-V)$.

In different samples of stars from Planet Search Programs, some
observers included a number of subgiants with a larger values of
magnitudes $M_V$ among the solar-type stars belonging to the
main-sequence. We try to exclude them from our analysis.

In Figure 2, one can see that stars have significantly different
values of $S$-index, which determines their chromospheric activity.
According to Isaacson et al. 2010, the mean levels of chromospheric
activity (corresponding to the uniform Mount Wilson $S$-index) for
stars of the spectral class F are higher than that of the G-stars.
On the other hand, for stars of K and M spectral classes, mean
levels of chromospheric activity are also higher than that of
G-stars.

We cannot see the close relationship of chromospheric activity of
stars from our samples versus the color index in Figure 2.

\begin{figure}[tbh!]
\centerline{
\includegraphics[width=140mm]{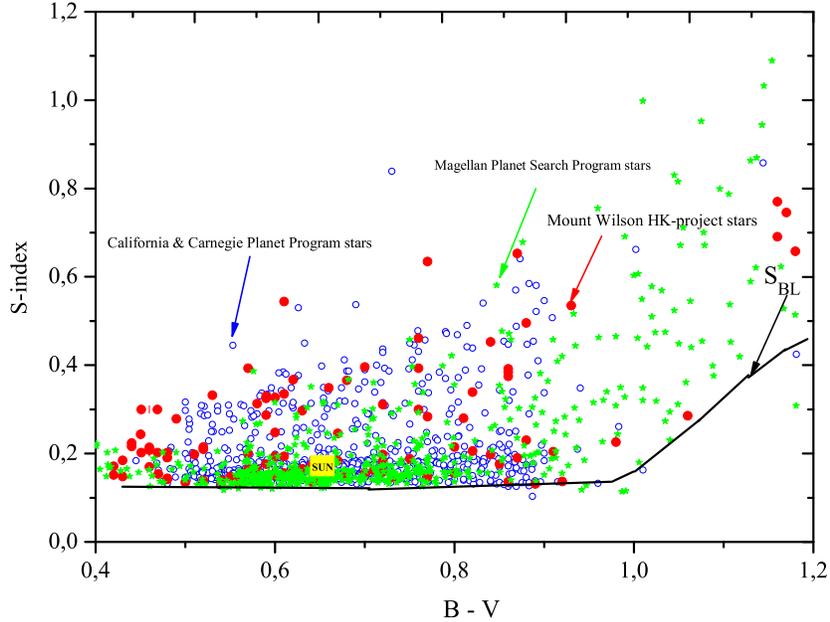}}
 \caption{Chromospheric activity of F, G, K and M stars from three observational Programs.}
 {\label{Fi:Fig2}}
\end{figure}

For the large statistical sample -- 2600 stars of the California
Planet Search Program -- the lower envelope of chromospheric
activity level $S_{BL}$ ($S$-index of Basic Level) was defined as a
function (polynomial fit) of $B-V$ for main-sequence stars over the
color range $0.4<B-V<1.6$, see Isaacson et al. 2010.

We show the $S_{BL}$ dependence in Figure 2. It is seen that the
Basic chromospheric activity Level $S_{BL}$  begins to rise when
$B-V>1$. Isaacson et al. 2010 believed that this increase of
$S_{BL}$ is due to decrease of continuum flux for redder stars:
$S$-value is defined as the ratio of $H$ and $K$ $CaII$ emission to
the nearby continuum.

It can be noted that the level of chromospheric activity of the Sun
is slightly below than the average level of chromospheric activity
of the stars belonging to the main sequence.

In solar-type stars of late spectral classes, X-rays are generated
by the magnetically confined plasma known as the corona (Vaiana et
al. 1981), which is heated by the stellar magnetic dynamo. The
observed decrease of the X-ray emission between pre-main sequence
young stars and older stars can be attributed to the rotational
spin-down of a star, driven by mass loss through a magnetized
stellar wind (Skumanich 1972).

\begin{figure}[tbh!]
\centerline{
\includegraphics[width=140mm]{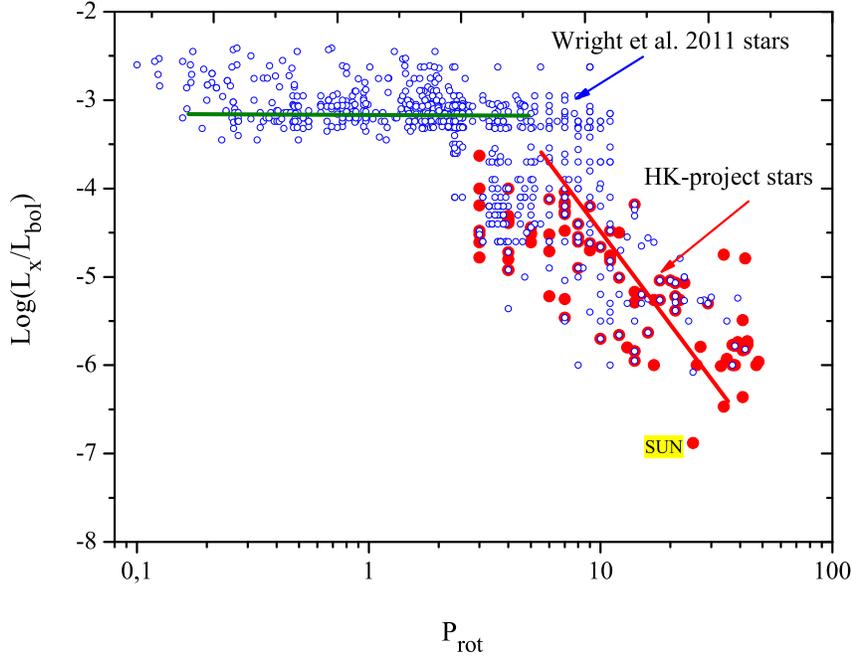}}
 \caption{X-ray to bolometric luminosity ratio plotted against rotation period.}
{\label{Fi:Fig3}}
\end{figure}

A relationship between stellar rotation and X-ray luminosity $L_X$
 was first described by Pallavicini et al. 1981.

 In Figure 3 we show the $L_X/L_{bol}$ versus $P_{rot}$ for
 stars from the catalog of stars, the details of which are
 presented in Wright et al. 2011, and for the stars of the HK-project.
In Write et al. 2011 there was done convertation of all X-ray
luminosities of the stars of a catalog of 824 stars to the ROSAT
0.1-2.4 keV energy band. The selected data of X-ray luminosities
$L_X/L_{bol}$ of 80 HK-project stars was also taken from the ROSAT
All-Sky Survey, see Bruevich et al. 2001. This sample also includes
the Sun using the values of $log L_X/L_{bol}=-6.24$ and
$P_{rot}=26.09$.
 Figure 3 demonstrates that there are two main regimes of coronal activity: a linear regime where
 activity increases with decreasing the rotation period, and a saturated
 regime where the X-rays luminosity ratio is constant with
 $log(L_X/L_{bol})=~-3.13$, see Wright et al. 2011.

 The HK-project stars
 which are not enough young and active are belong to the stars of linear
 regime in Figure 3. We can see that the Sun confirms its unique place among
 solar-type stars: it has almost the lowest
 level of X-rays luminosity among all the stars.

Below in Figure 7b we also show the place of the solar coronal
activity among the coronal activity of stars of the HK-project. It
can be noted the fact that the Sun is at the place with absolutely
lowest level of coronal activity among solar-type stars.

It should be noted that solar photometric radiation changed very
little in the activity cycle, less than 0.1 \%.
 The monitoring of the photometric and chromospheric $H$ and $K$ $Ca II$
 emission data series of stars similar to the Sun in age and average activity
level showed that there is an empirical correlation between the
average stellar chromospheric activity level and the photometric
variability. In general, more active stars show larger photometric
variability. The Sun is significantly less variable that indicates
by the empirical relationship, see Shapiro et al. 2013. It was found
that on a long time scale the position of the Sun on the diagram of
photometric variability versus chromospheric activity changers is
not constant in time. So Shapiro et al. 2013 suggested that the
temporal mean solar variability might be in agreement with the
stellar data.

But at present we can see that the Sun confirms its unique place
among solar-type stars: its photometric variability is unusually
small. This is confirmed in Lockwood et al. 2007: the photometric
observations at the Lowell observatory of 33 stars of the HK-project
revealed the fact that the
 photometric variability of the Sun during the cycle of magnetic activity
 is much less than the photometric variability of the other HK-project
 stars.

\vskip12pt
\section{Observations of HK-project stars}
\vskip12pt

It can be noted that among the databases of observations of
solar-type stars with known values of $S$-index the sample of stars
of HK-project was selected most carefully in order to study the
stars which are analogs of the Sun.

Moreover, unlike different Planet Search Programs of observations of
solar-type stars, the Mount Wilson Program was specifically
developed for the study of solar-type cyclical activity of the
main-sequence F, G and K-stars (single) which are the closest to the
"young Sun" and "old Sun".

The duration of observations more than 40 years during the
HK-project has allowed to detect and explore the cyclical activity
of the stars, similar to 11-yr cyclical activity of the Sun. First
O. Wilson began this program in 1965. He attached great importance
to the long-standing systematic observations of cycles in the stars.
The fluxes in passbands 0.1 nm wide and centered on the $CaII$ $H$
and $K$ emission cores have been monitored in 111 star of spectral
type F2-K5 on or near main sequence on the Hertzsprung-Russell
diagram (Baliunas et al. 1995; Radick et al. 1998; Lockwood et al.
2007).

Mount Wilson $S$-index is the relationship of radiation fluxes in
the centers of emission lines $H$ and $K$ (396,8 nm and 393,4 nm) to
radiation fluxes in the near-continuum (400,1 and 390,1 nm) - is
considered now as a sensitive indicator of the chromospheric
activity of the Sun and the stars.

For HK-project the stars were carefully chosen according to those
physical parameters, which were  most close to the Sun: cold, single
stars-dwarfs, belonging to the main sequence. Close binary systems
are excluded.

The results of the joint observations of the HK-project  radiation
fluxes and periods of rotation gave the opportunity for the first
time in stellar astrophysics to detect the rotational modulation of
the observed fluxes (Noyes et al. 1984). This meant that on the
surface of the star there are inhomogeneities those were living and
evolving in several periods of rotation of the stars around its
axis. In addition, the evolution of the periods of rotation of the
stars in time clearly pointed to the fact of existence of the star's
differential rotations similar to the Sun's differential rotations.

The authors of the HK-project with the help of frequency analysis of
the 40-year observations have discovered that
 the periods of star's 11-yr cyclic activity vary
little in size for the same star (Baliunas et al. 1995; Lockwood et
al. 2007). The durations of cycles vary from 7 to 20 years for
different stars. The stars with cycles represent about 30 \% of the
total number of studied stars.

\vskip12pt
\section{Cyclic activity of HK-project stars. From periodogram to wavelet analysis.}
\vskip12pt

The evolution of active regions on the star on a time scale of about
10 years determines the cyclic activity similar to the Sun.

For 111 HK-project stars the, periodograms were computed for each
stellar record in order to search for activity cycles, see (Baliunas
et al. 1995). The significance of the height of the tallest peak of
the periodogram was estimated by the false alarm probability (FAP)
function, see Scargle (1982). The stars with cycles  were classified
as follows: if for the calculated $P_{cyc} \pm \Delta P$ the FAP
function $\leqslant 10^{-9}$ then this star is of "Excellent" class
($P_{cyc} $ is the period of the cycle). If $ 10^{-9} \leqslant FAP
\leqslant 10^{-5} $ then this star is of "Good" class. If $ 10^{-5}
\leqslant FAP \leqslant 10^{-2} $ then this star is of "Fair" class.
If $ 10^{-2} \leqslant FAP \leqslant 10^{-1} $ then this star is of
"Poor" class.

About 50 of 111  stars  were identified with varying degrees of
reliability as the stars with regular cycles which periods $P_{cyc}
$ which are varied from 7 to 20 years.

\begin{figure}[h!]
   \centerline{\hspace*{0.005\textwidth}
               \includegraphics[width=70mm]{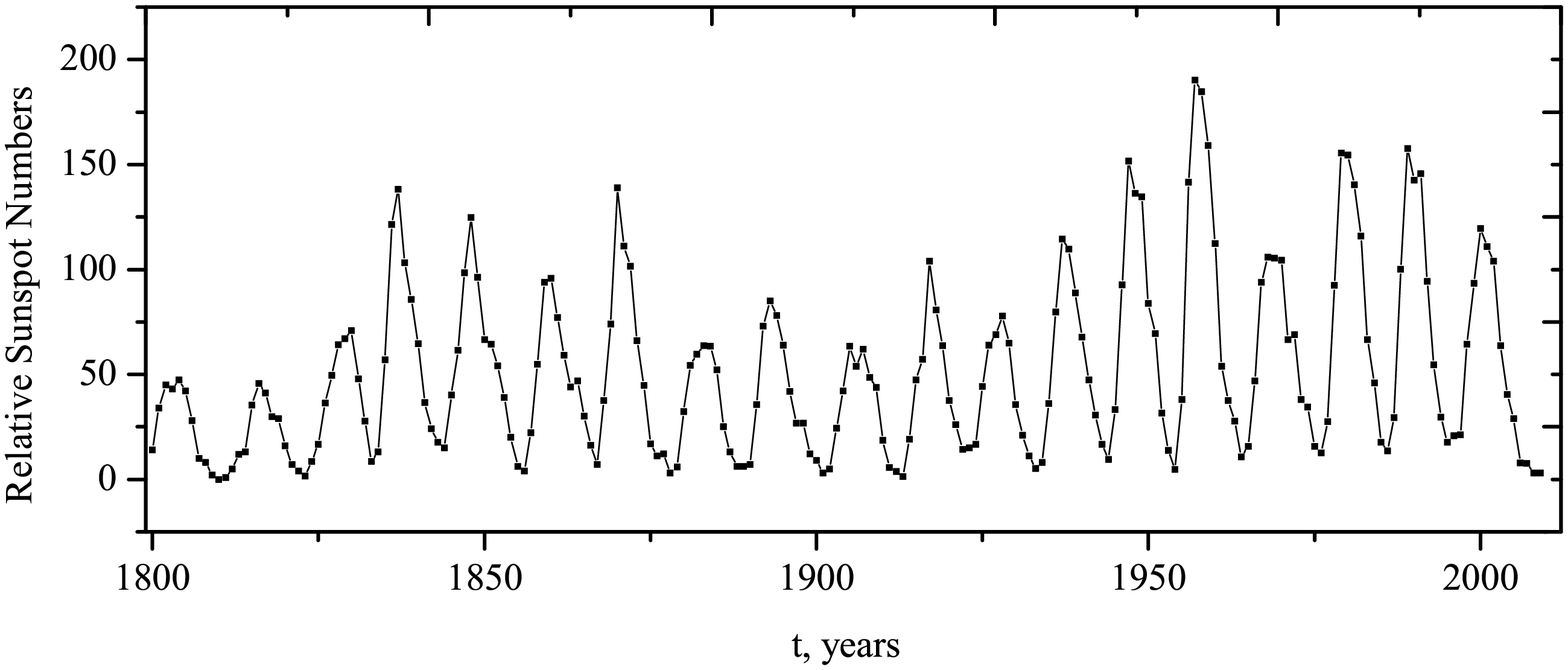}
               \includegraphics[width=70mm]{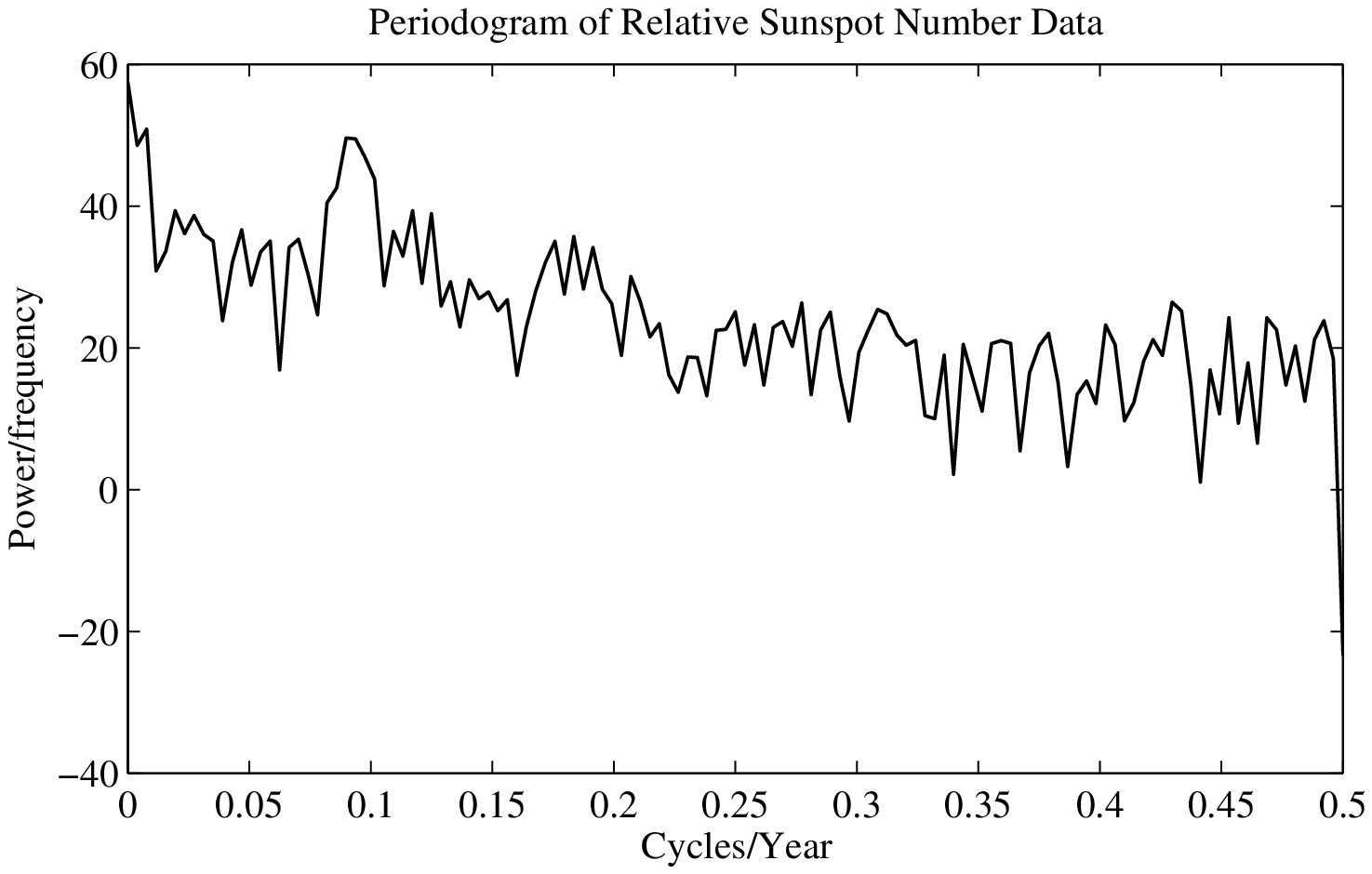}
              }

\caption{Yearly averaged relative sunspot numbers, observations from
1800 to 2009 -- at left; the periodogram of the yearly averaged
relative sunspot numbers for 1800 to 2009 observations -- at right.}
 {\label{Fi:Fig4}}
\end{figure}

\begin{figure}[tbh!]
\centerline{
\includegraphics[width=120mm]{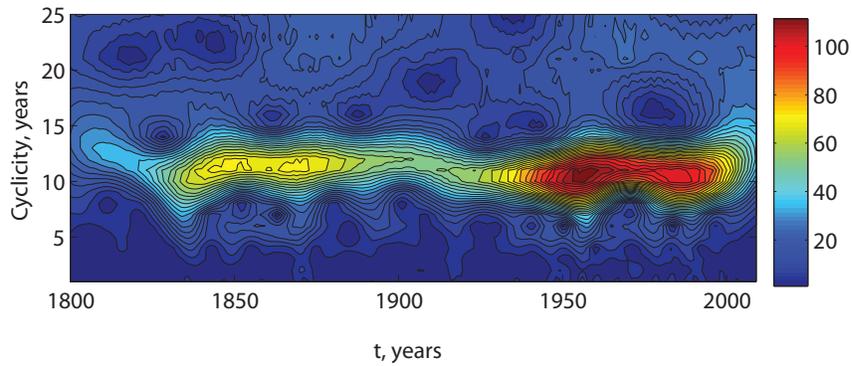}}
 \caption{Wavelet analysis  of the yearly averaged relative sunspot numbers. Complex Morlet wavelet 1.5-1.
Observations from 1800 to 2009. }
 {\label{Fi:Fig5}}
\end{figure}

In (Baliunas et al. 1995; Radick et al. 1998; Lockwood et al. 2007)
the regular chromospheric cyclical activity of  HK-project
solar-type stars were studied through the analysis of the power
spectral density with the Scargle's periodogram method (Scargle
1982).
%Scargle 1982 pointed out that the detection of a periodic signal hidden by a
%noise is frequently a goal in astronomical data analysis. So, the
Periods of HK-project stars activity cycles similar to 11-yr solar
activity cycle were calculated. The significance of the height of
the tallest peak of the periodogram was estimated by the false alarm
probability (FAP) function. In the case when the calculated $P_{cyc}
\pm \Delta P$ the FAP function $\leqslant 10^{-9}$ this star is of
"Excellent" class. Among these 50 stars with cycles they have found
only 12 stars and the Sun to be characterized by the cyclic activity
of the "Excellent" class.

We illustrate the method of cyclic period calculation with Scargle's
periodogram technique on the example of the Sun. We obtained the
periodogram of the  yearly averaged relative sunspot number (or Wolf
number) for observations conducted in 1800 -- 2009.

In Figure 4 (left part) we present the relative sunspot number
yearly averaged data set. It is known that the direct observations
of sunspots were made only since 1850, and from 1800 to 1850 years
the sunspots data were taken from indirect estimates. This fact, as
we will see below in Figure 5, affects the quality of the wavelet
analysis from 1800 to 1850 years.

The Scargle's periodogram of the relative sunspot number for 1800 -
2009 data set is presented in Figure 4 (right part). This
periodogram shows that there is a peak at approximately 0.1
cycles/year, which indicates a period of approximately 10 years.

Our illustration  of the Scargle's  periodogram method shows that,
unfortunately, this method allows us only to define a fixed set of
the main frequencies (that determines the presence of significant
periodicities in the series of observations). In the case where the
values of periods change significantly during the interval of
observations, the accuracy of determination of periods becomes
worse. It is also impossible to obtain information about the
evolution of the periodicity in time.

In Figure 5 we show the results of wavelet analysis of the same
sunspot number yearly averaged data set.

Figure 5 confirms the known fact that the periods of the solar
activity cycle is about 11-yr in the XIX century and about 10 yr in
the XX century. It is seen that the quality of the wavelet pictures
in the first half of the XIX century is much worse than later in the
XX century when the era of direct observation began.

According to different solar observations, the mean value of the
period of solar activity cycle in the twentieth century is about
10.2 years. It is also known that the abnormally long 23-rd cycle of
solar activity ended in 2009 and lasted about 12.5 years. All of
this is shown in relative sunspot numbers wavelet picture in Figure
5. Thus, we can see that the value of period of the main cycle of
solar activity for the past 150 years is not constant and varies by
15-20 \%.

In Kollath and Olah 2009 it has been  tested and used different
methods, such as short-term Fourier transform, wavelet, and
generalized time-frequency distributions, for analyzing temporal
variations in timescales of long-term observational data which have
information on the magnetic cycles of active stars and that of the
Sun.

Wavelet analysis is becoming a common tool in present time for the
analysis of localized variations of power within a time series. By
decomposing a time series into time-frequency space, one is able to
determine  the dominant modes of variability and how those modes
vary in time. The choice of wavelet is dictated by the signal or
image characteristics and the nature of the application.
Understanding the properties of the wavelet analysis and synthesis,
you can choose a mother wavelet function that is optimized for your
application. Thus, we choose a complex Morlet wavelet which depends
on two parameters: a bandwidth parameter and a wavelet center
frequency.

The multi-scale evolution in the solar activity was found in Kollath
and Olah 2009 with used of wavelet and generalized time-frequency
analysis. The observed features in the time-frequency history of the
Sun were analyzed with wavelet methods. The time-frequency analysis
of  multi-decadal variability of the solar Schwabe (11-yr) and
Gleissberg (secular)  cycles during the last 250 years from Sunspot
Number records showed that one cycle (Schwabe) varies between
limits, while the longer one (Gleissberg) continually increases. In
Kollath and Olah 2009 by analogy from the analysis of the longer
solar record, the presence of a long-term trend may suggest an
increasing or decreasing of multi-decadal cycle that is presently
unresolved in the stellar records of short duration.

In Olah et al. 2009 the study of the  time variations of the cycles
of  20 active stars based on decades-long photometric or
spectroscopic observations with method of time-frequency analysis
was done. They found that the cycles in the sun-like stars show
systematic changes, the same phenomenon we can see in the cycles of
the Sun.

Olah et al. 2009 found that fifteen stars definitely show multiple
cycles, the records of the rest are too short to verify a timescale
for a second cycle. For 6 HK-project stars: HD 131156A, HD 131156B,
HD 100180, HD 201092, HD 201091 and HD 95735 the multiple cycles
were detected with used of wavelet-analysis.

HD 131156A shows variability on two time scales: the shorter cycle
is about 5.5-yr, a longer-period variability is about 11 yr. For HD
131156B only one long-term periodicity has been determined. For HD
100180 the variable cycle of 13.7-yr appears in the beginning of the
record; the period decreases to 8.6-yr by the end of the record. The
results in the beginning of the dataset are similar to those found
by Baliunas et al. 1995, who found two cycles, which are equal to
3.56 and 12.9-yr. The record for HD 201092 also exhibits two
activity cycles: one is equal to 4.7-yr, the other has a timescale
of  10-13 years. The main cycle seen in the record of HD 201091 has
a mean length of 6.7-yr, which slowly changes between 6.2 and
7.2-yr, a shorter, significant cycle is found in the first half of
the record with a characteristic timescale of 3.6-yr. The stronger
cycle of HD 95735 is 3.9-yr, which is slightly shorter (3.4-yr),  a
longer, 11-yr cycle is also present with a smaller amplitude.

We showed that complex Morlet wavelet 1.5 - 1 can most accurately
determine the dominant cyclicity as well as its evolution in time in
solar data sets at different wavelengths and spectral intervals
(Bruevich et. al 2013).

We have applied the wavelet analysis for partially available data in
form of pictures (the curves of variation of $S$-index with time in
Baliunas et al. 1995 and Lockwood et al. 2007) to 5 HK-project stars
with cyclic activity of the "Excellent" class: HD 10476,  HD 81809,
HD 103095, HD 152391, HD 160346 and to the star HD 185144 with no
cyclicity.

We used the complex Morlet wavelet 1.5 - 1 which can most accurately
determine the dominant cyclicity as well as its evolution in time in
solar data sets at different wavelengths and spectral intervals
(Bruevich et. al 2013).

 So for these 6 stars we have used partially
available observation data of $S$-index for 1965-1992 observation
sets (Baliunas et al. 1995) and for 1985-2002 observations (Lockwood
et al. 2007). We used the detailed graphical time dependencies of
$S$-index, each point of the record of observations, which we
processed in this paper using wavelet analysis technique,
corresponds to three months averaged values of $S$-index.

We hoped that wavelet analysis can help us to study the temporal
evolution of chromospheric activity cycles of the stars. Tree-month
averaging also helps us to avoid the modulation of observational
data by star's rotations similarly to the case of the Sun.

In Figure 6 we present our results of the cycles of 6 stars:

HD 81809 has a mean length of 8.2-yr, which slowly changes between
8.3-yr in the first half of the record and 8.1-yr in the middle and
the end of the record while Baliunas et al. 1995 found 8.17-yr.

HD 103095 has a mean length of 7.2-yr, which slowly changes between
7.3-yr in the first half of the record, 7.0-yr in the middle and
7.2-yr in the end of the record while Baliunas et al. 1995 found
7.3-yr.

HD 152391 has a mean length of 10.8-yr, which slowly changes between
11.0-yr in the first half of the record and 10.0-yr in the end of
the record while Baliunas et al. 1995 found 10.9-yr.

HD 160346 has a mean length of 7.0-yr which is not changes during
the record in agreement with Baliunas et al. 1995 estimated 7.0-yr.

HD 10476 has a mean length of 10.0-yr in the first half of the
record, then the length sharply changes to 14-yr, while Baliunas et
al. 1995 found 9.6-yr. After changing of high amplitude cycle's
period from 10-yr to 14-yr in 1987 the low amplitude cycle remained
exist at 10.0-yr period - we can see two activity cycles. Baliunas
et al. 1995 estimated HD 10476 cycle as 9.6-yr.

HD 185144 has a mean length of 7-yr which changes between 8-yr in
the first half of the record and 6-yr in the end of the record while
Baliunas et al. 1995 haven't found the well-pronounced cycle.

In Olah et al. 2009 the multiple cycles were found for the stars HD
13115A, HD 131156B,  HD 93735 for which in Baliunas et al. 1995
haven't been found any cycles; for the stars of "Excellent" class HD
201091 and HD 201092 the cycle's periods found in Baliunas et al.
1995 were confirmed and also the shorter cycles (similar to solar
quasi-biennial) were determined.

Olah et al. 2009 have concluded that all the stars from their
pattern of cool main-sequence stars have the cycles and most of the
cycle lengths change systematically. But the stars of "Excellent"
class have relatively constant cycle lengths - for these stars the
cycle's periods calculated in Baliunas et al. 1995 and cycle's
periods found with use of wavelet analysis are the same.

\begin{figure}[h!]
   \centerline{\hspace*{0.0005\textwidth}
               \includegraphics[width=70mm]{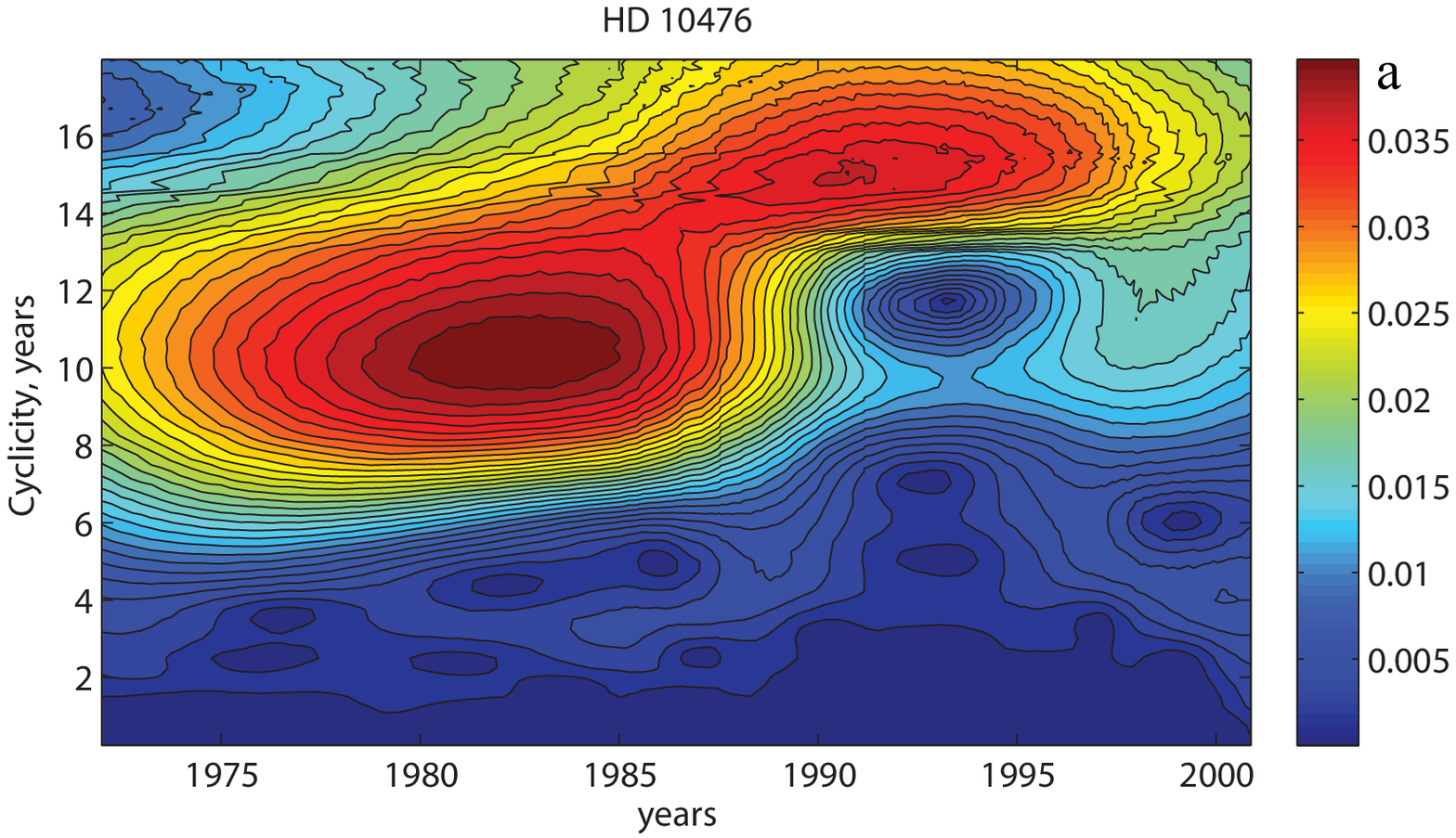}
               \includegraphics[width=70mm]{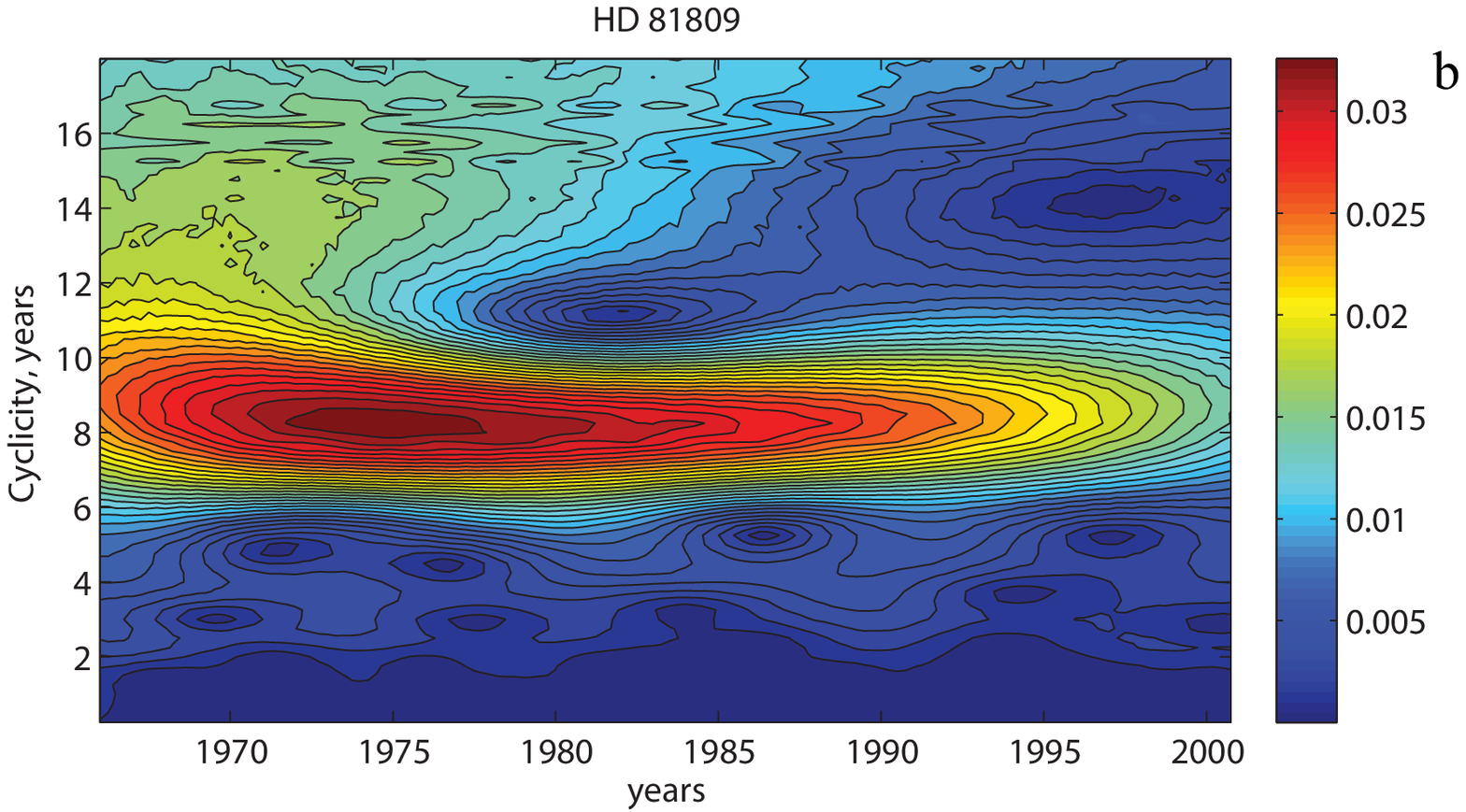}
              }
   \centerline{\hspace*{0.015\textwidth}
               \includegraphics[width=70mm]{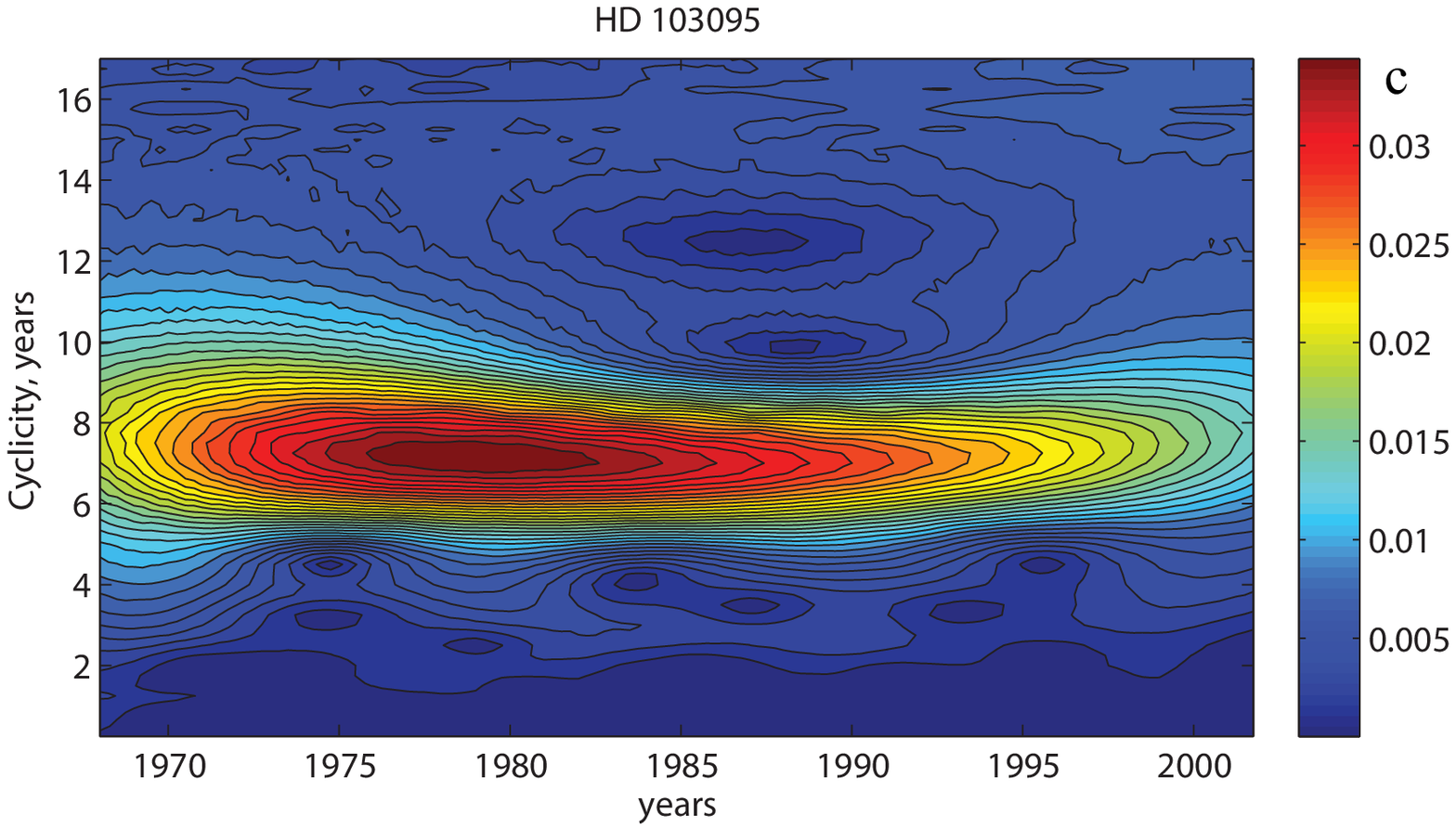}
               \includegraphics[width=70mm]{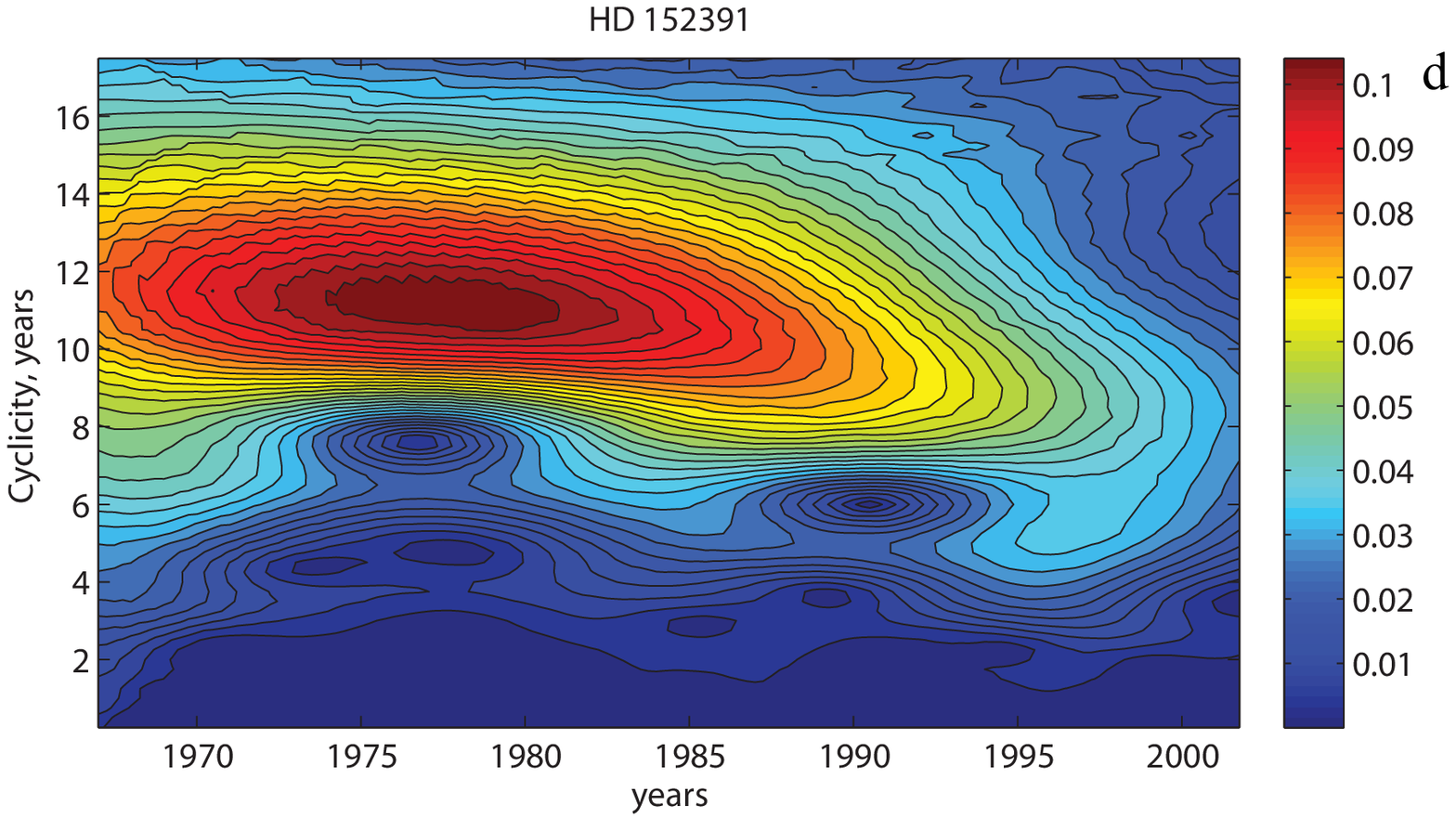}
              }
   \centerline{\hspace*{0.015\textwidth}
               \includegraphics[width=70mm]{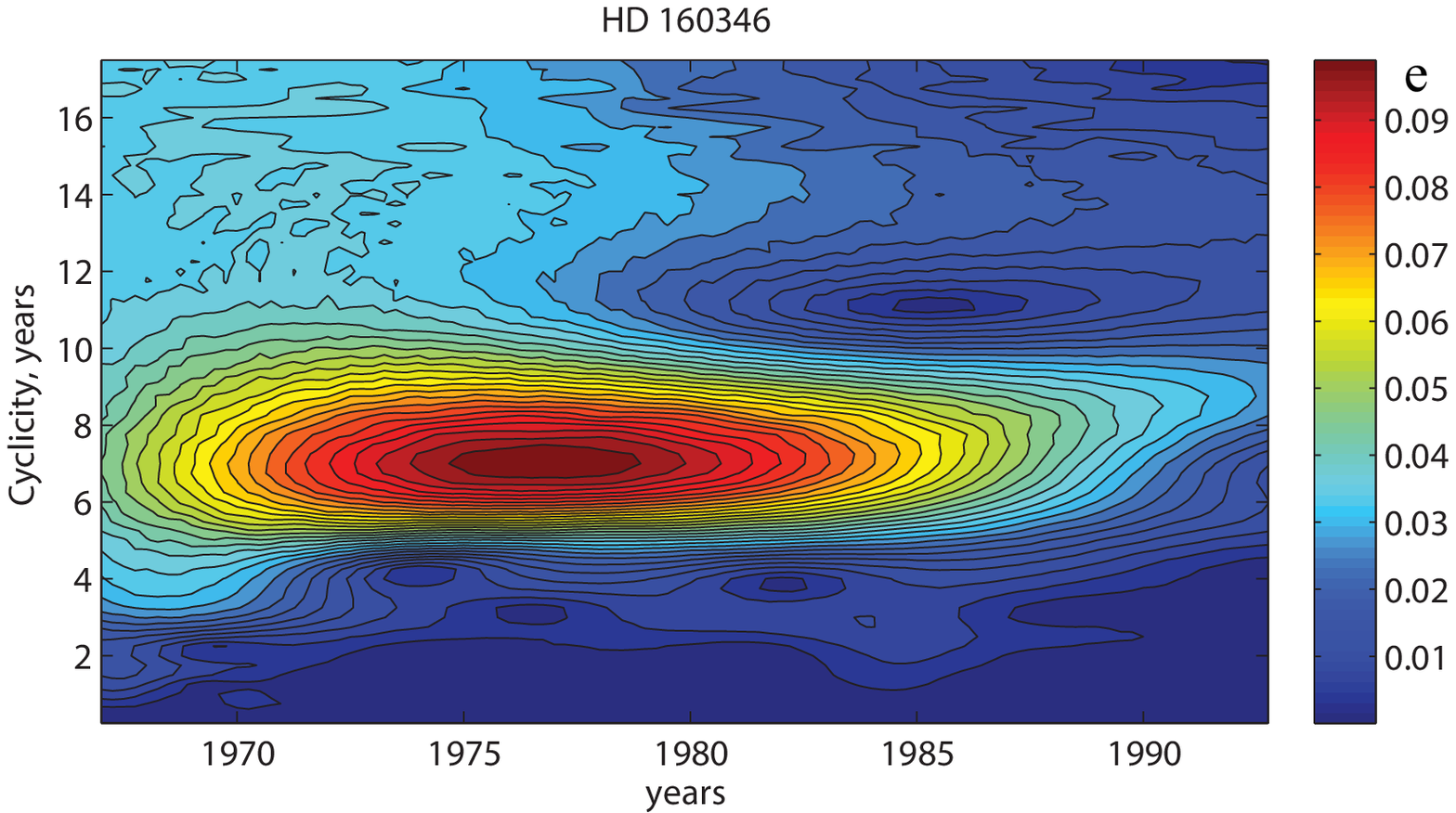}
               \includegraphics[width=70mm]{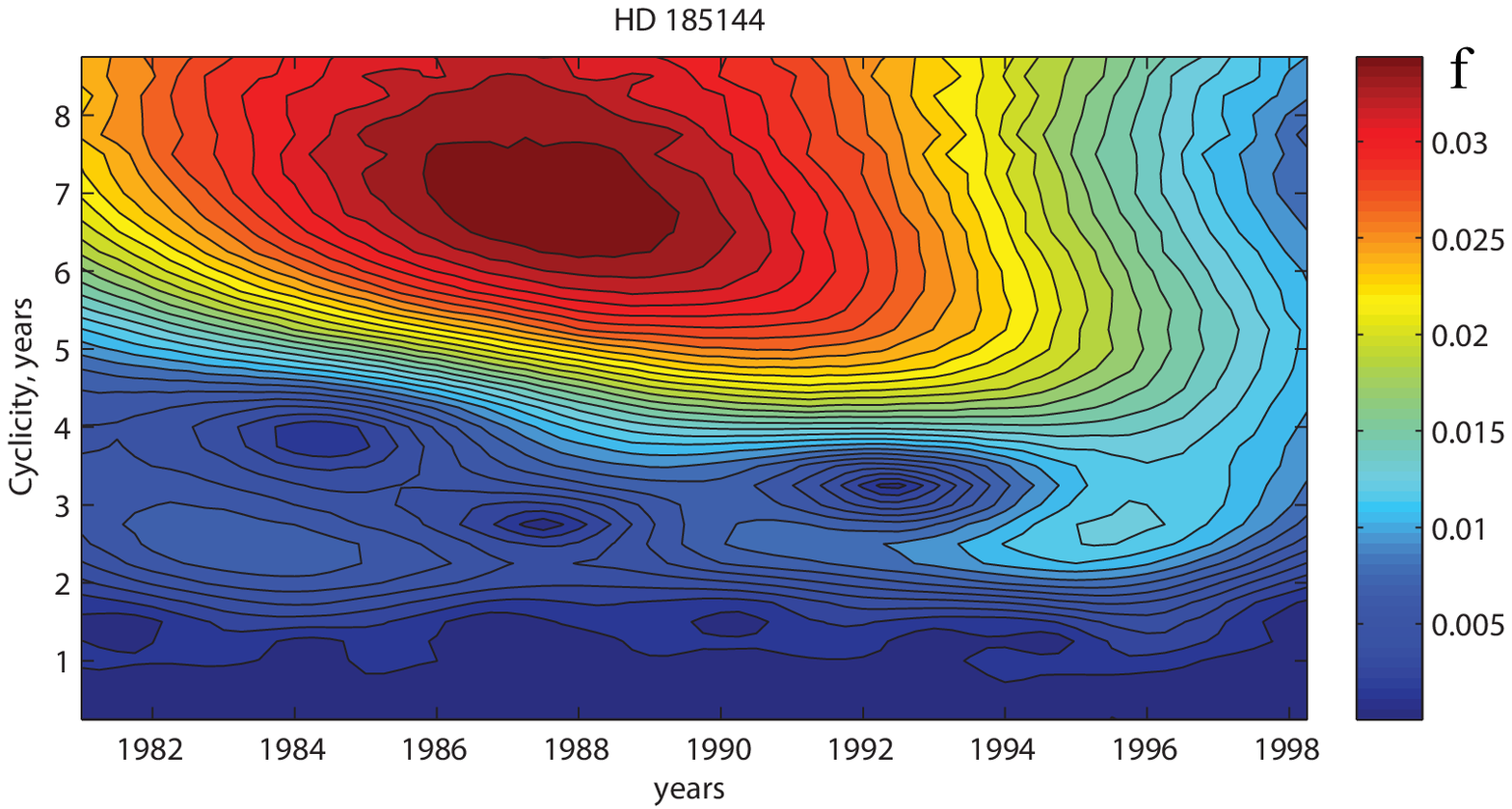}
               }

\caption{Wavelet analysis  of HK-project stars. Complex Morlet
wavelet 1.5-1. Observations from 1969 to 2002: (a) HD 10476, (b) HD
81809, (c) HD 103095, (d) HD 152391, (e) HD 160346, (f) HD 185144.}

   \label{F-6panels}
   \end{figure}

The long-term behavior of the sunspot group numbers have been
analyzed using wavelet technique by Frick et al. 1997 who plotted
the changes of the Schwabe cycle (length and strength) and studied
the grand minima. The temporal evolution of the Gleissberg cycle can
also be seen in the time-frequency distribution of the solar data.
According to Frick et al. 1997 the Gleissberg cycle is as variable
as the Schwabe cycle. It has two higher amplitude occurrences: first
around 1800 (during the Dalton minimum), and then around 1950. They
found very interesting fact - the continuous decrease in the
frequency (increase of period) of Gleissberg cycle. While near 1750
the cycle length was about 50 yr, it lengthened to approximately 130
yr by 1950.

In the late of XX century some of solar physicists began to examine
with different methods the variations of relative sunspot numbers
not only in high amplitude 11-yr Schwabe cycle but in low amplitude
cycles approximately equal to half (5.5-yr) and fourth
 (quasi-biennial) parts of period of the main 11-yr cycle, see (Vitinsky et al. 1986).
The periods of the quasi-biennial cycles vary considerably within
one 11-yr cycle, decreasing from 3.5 to 2 yrs, and this fact
complicates the study of such periodicity using the method of
periodogram estimates.

Using the methods of frequency analysis of signals  the
quasi-biennial cycles have been  studied not only for the relative
sunspot number, but also for 10.7 cm solar radio emission and for
some other indices of solar activity ((Bruevich et al. 2013,
Bruevich and Yakunina 2015). It was also shown that the cyclicity on
the quasi-biennial time scale takes place often among the stars with
11-yr cyclicity, see Bruevich and Kononovich 2011.

The cyclicity similar to the solar quasi-biennial was also detected
for the solar-type stars from the direct observations.
 In Morgenthaler et al. 2011 the results of direct observation of magnetic cycles of
19 solar-type stars of F, G, K spectral classes within 4 years were
presented. The stars of this sample are characterized by masses
between 0.6 and 1.4 solar mass and by rotation periods between 3.4
and 43 days. Observations were made using NARVAL spectropolarimeter
(Pic du Midi, France) between 2007 and 2011. It was shown that for
the stars of this sample $\tau$ Boo and HD 78366 (the same of the
Mount Wilson HK-project) the cycle lengths derived by chromospheric
activity (Baliunas et al. 1995) seem to be longer than those derived
by spectropolarimetry observations of Morgenthaler et al. 2011. They
suggest that this apparent discrepancy may be linked to the
different temporal sampling inherent to the two approaches, so that
the sampling adopted at Mount Wilson may not be sufficiently tight
to unveil short activity cycles. They hope that future observations
of Pic du Midi stellar sample will allow them to investigate longer
time scales of the stellar magnetic evolution.

For the solar-type F, G and K stars according to {\it Kepler}
observations, "shorter" chromosphere cycles with periods of about
two years have also been found, see Metcalfe et al. 2010, Garcia et
al. 2010.

We assume that precisely these quasi-biennial cycles were identified
in Morgenthaler et al. 2011: $\tau$ Boo and HD 78366 are the same of
the HK-project, these stars have the cycles similar to the
quasi-biennial solar cycles with  periods of a quarter of the
duration of the periods defined in Baliunas et al. 1995.

Note, that in case of the Sun, the amplitude of variations of the
radiation in quasi-biennial cycles is substantially less than the
amplitude  of variations in main 11-yr cycle. We believe that this
fact is also true for all solar-type stars of the HK-project and in
the same way for $\tau$ Boo and HD 78366.

The quasi-biennial cycles cannot be detected with the Scargle's
periodogram method. But methods of spectropolarimetry from
(Morgenthaler et al. 2011) allowed detecting the cycles with 2 and
3-yr periods. Thus spectropolarimetry is more accurate method for
detection of cycles with different periods and with low amplitudes
of variations.

So, the need for wavelet analysis of HK-project observational data
is dictated also by the fact that the application of wavelet method
to these observations will help: (1) to find the cyclicities with
periods equal to half and a quarter from the main high amplitude
cyclicity; (2) to clarify the periods of the high amplitude cycles
and to follow their evolution in time; (3) to find still other stars
with cycles for which the cycles were not determined using the
method of periodogram due to strong variations of the period as in
case of HD 185144.

The analysis of cyclic activity of solar-type stars using Scargle's
periodogram method in (Baliunas et al. 1995) and wavelet analysis
simultaneously showed that the selection of stars into classes
according to the quality of their cycles ("Excellent", "Good",
"Fair" and "Poor") is very important moment in the study of stellar
cycles.

Wavelet analysis helped us to understand why the stars of "Fair" and
"Poor" classes are differ from the stars of "Excellent" and "Good"
classes: the main peak on their periodograms is greatly expanded due
to strong variations of the cycle's duration.

As it turned out, the differentiation of stars with cycles onto
classes "Excellent" , "Good", "Fair" and "Poor" is very important:
the stars with the stable cycles "Excellent" and "Good" and  the
stars with the unstable cycles "Fair" and "Poor"   relate to
different groups in the graphs of dependencies $S$ versus $(B-V)$,
$logL_X $ versus $(B-V)$,  $P_{cyc}$ versus Age, see Figure 7 and
Figure 8 below.

\vskip12pt
\section{Chromospheric and coronal activity of HK project stars of different spectral classes with cycles}
\vskip12pt

Processes, that determine complex phenomena of stellar activity and
covering practically the whole atmosphere from the photosphere to
the corona, occur differently among solar-type stars belonging to
different spectral classes.

The Mount Wilson HK project observational data allow us to study the
solar-type cyclic activity of stars simultaneously with their
chromospheric and coronal activity. The selected data of X-ray
luminosities $log L_X$ of 80 HK-project stars is taken from the
ROSAT All-Sky Survey, see Bruevich et al. 2001.

\begin{figure}[h!]
  \centerline{\hspace*{0.0005\textwidth}
               \includegraphics[width=90mm]{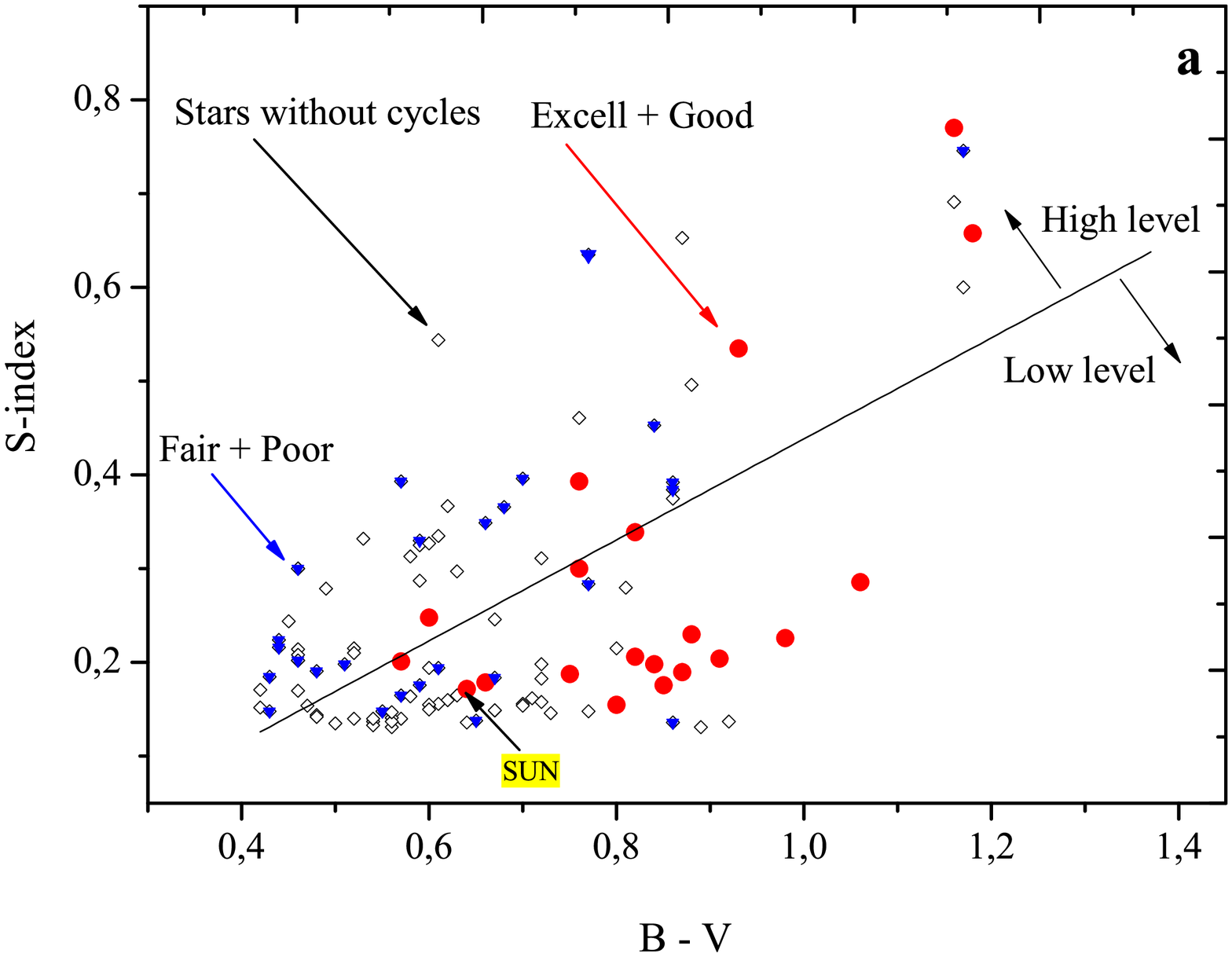}
               \includegraphics[width=90mm]{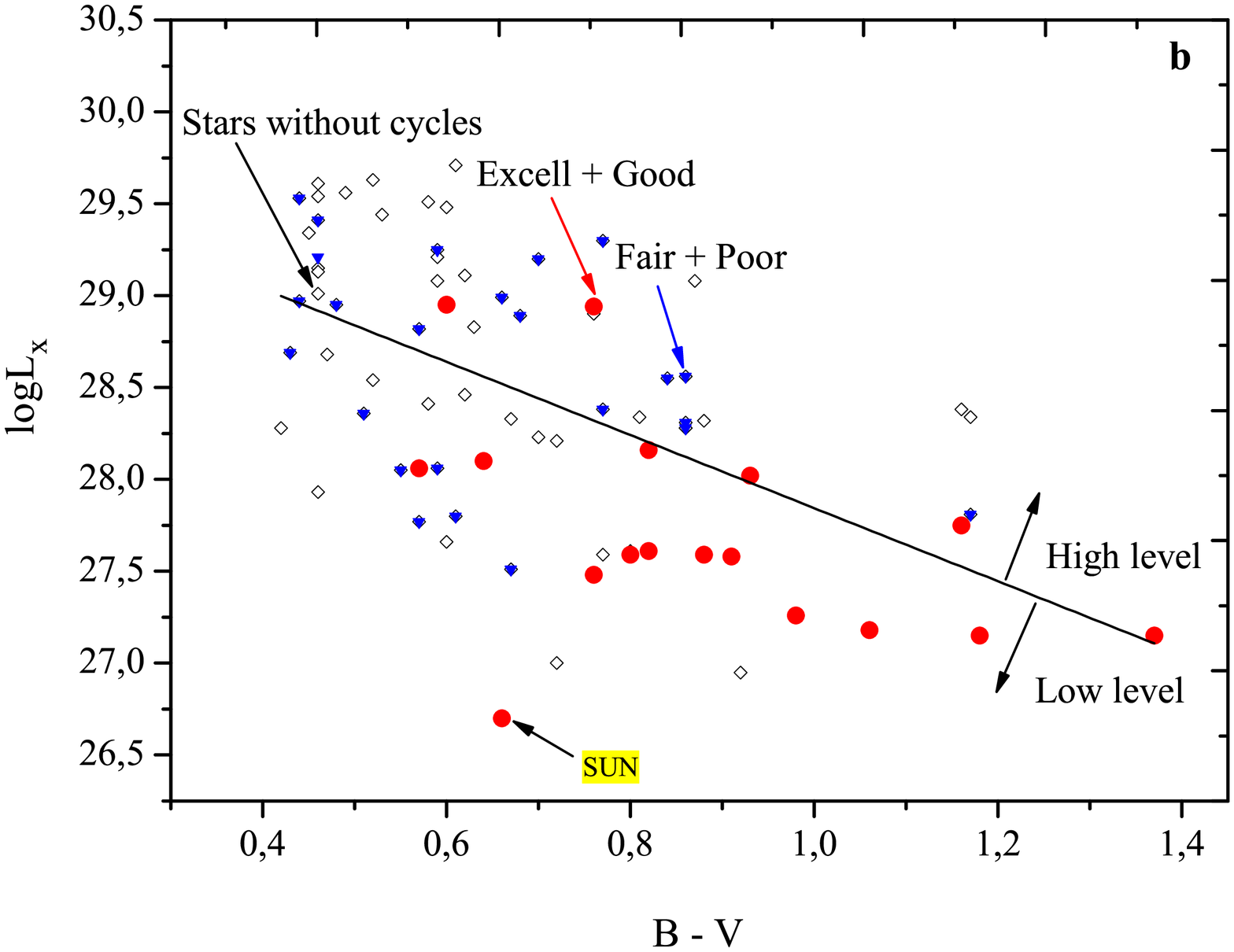}
              }
\vskip12pt

 \centerline{\hspace*{0.15\textwidth}
               \includegraphics[width=90mm]{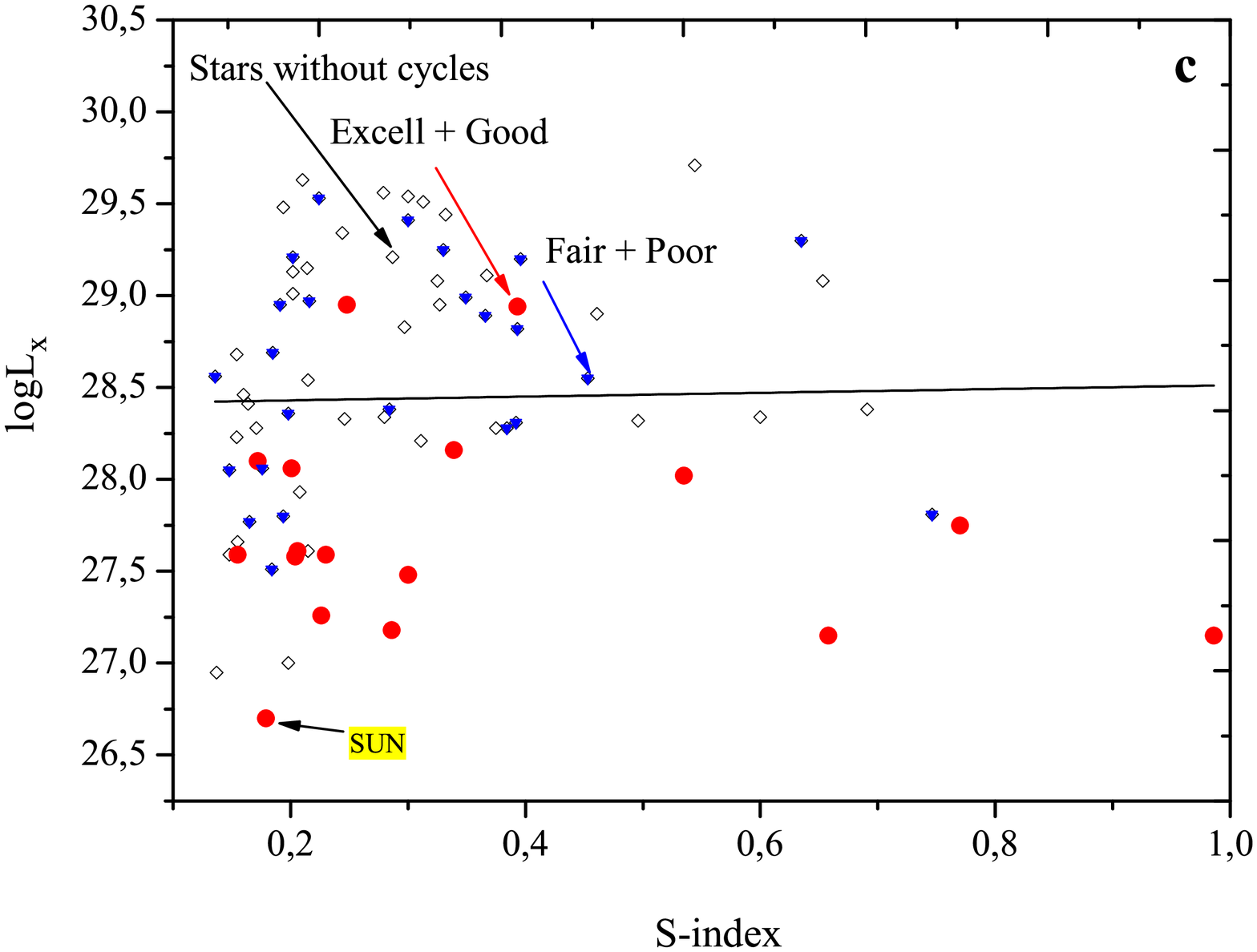}}
\caption{HK-project stars. Observations from 1969 to 1994 yrs. (a)
$S$ versus $(B-V)$, (b) $log L_X$ versus $(B-V)$, (c)  $log L_X$
versus $S$.}

   \label{F-3panels}
   \end{figure}

As noted earlier (Baliunas et al. 1995), the average chromospheric
activity of the stars, or rather the value of $log S$ varies
(increases) with the increase of the color index $(B-V)$, see Figure
2, Figure 7a.

Our linear regression analysis of HK-project stars showed that there
is a relation which is described by the following formula:

    $$  S = ~- ~0.10 + 0.530  \cdot (B-V)   ~ ~ ~  (1)  $$

  Let us denote the right part of the relation (1) as $F(B-V)$.

We consider the stars which have $S > F(B-V)$ to be characterized by
the high level of chromospheric activity, and stars with $ S
\leqslant F(B-V)~- $  by the low level of chromospheric activity,
see  Figure 7a.

Next, we have analyzed all 110 stars from the HK-project and the Sun
to determine which kind of the level of chromospheric activity
corresponds to one or another star. We will consider these results
further in the comparative analysis of stars of different spectral
classes, see Table 1.

For 80 stars, the coronal radiation of which we know from the ROSAT
data, we also do linear regression analysis and obtain the following
relationship between X-ray luminosity, normalized to the bolometric
luminosity, and color index $(B-V)$:

    $$ log L_X = ~ 29.83 - 1.99  \cdot (B-V)    ~ ~ ~  (2)  $$

Let us denote the right hand side of the relation (2) as  $P(B-V)$.
By analogy with the analysis of the chromospheric activity of stars,
we consider the stars with $ log L_X  > P(B-V)$ to be characterized
by the high level of coronal activity, and stars with $log L_X
\leqslant P(B-V) $ --- by the low level of coronal activity, see
Figure 7b.

As noted above, in the case of chromospheric activity a direct
correlation takes place: with the increase of the color index
$(B-V)$ the average value of chromospheric activity ($ S $)~of stars
increases. But in the case of X-ray radiation of stars from $(B-V)$
the inverse correlation takes place: with the increase of the color
index $(B-V)$, the average value $ log L_X$ decreases, see Figure
7a, 7b.

Note that most of stars, characterized by increased chromospheric
activity, have also increased coronal activity. About 15 \% of
stars, including the Sun, are characterized by coronal activity that
is significantly lower than the value which should correspond to its
chromospheric activity, see Figure 7c.

Figures 7a, 7b demonstrate that stars with cycles of "Excellent" and
"Good" classes are mostly characterized by low level of
chromospheric and coronal activity (about 70 \%), as opposed to
stars with cycles of "Fair" and "Poor" classes which are mostly
characterized by high level of chromospheric and coronal activity
(about 75 \% ).

The existence or absence of a pronounced cyclicity, as well as the
quality of the identified cycles (belonging to classes "Excellent",
"Good", "Fair", "Poor"), for F, G and K stars varies significantly.

Bruevich et al. (2001) noted the difference between the stars of
"Excellent", "Good", "Fair" and "Poor" classes from a position of
presence and degree of development of under photospheric convective
zones of stars of different $(B-V)$.

Thus, we can note here (illustrated below in Table 1) that the
quality of chromospheric activity cycles (the ratio of the total
number of stars belonging to classes with a well-defined cyclicity
"Excellent" + "Good" to the number of stars with less than a certain
cyclicity "Fair" + "Poor") essentially differs for stars of
different spectral classes F, G and K.

\begin{table}
\caption{Comparative analysis of cycles of stars and the quality of
their cyclicities for stars of different spectral classes.}
\vskip12pt
\begin{tabular}{clclclclclclclcl}

\hline \hline
                      &      &     &      \\
  Interval of spectral classes &    F2~-F9    &    G2~-G9 ~  &      K0~-K7   \\
\hline
                      &      &     &      \\
 $\Delta (B-V) $ & 0.42~-0.56   &0.57~-0.87~ &0.88~-1.37 \\
\hline
Total number of stars &      &     &      \\
in spectral interval    &39&44  &27      \\
\hline
   Number of stars with &      &     &      \\
   known values  $L_X$ & 27 & 29   & 24         \\
\hline
Relative number of stars  with   &      &     &      \\
  increased coronal activity    & 60\% &  ~48\% ~      &   ~41\%         \\
\hline
 Relative number of stars  with &      &     &      \\
 increased chromospheric activity    & 56\% &  ~39\% ~      &   ~60\%     \\
\hline
 relative number of stars with &      &     &      \\
 chromospheric activity cycles   & 25\% &  ~40\% ~      &   ~72\%         \\
\hline

Quality of chromospheric cycles   &      &     &      \\
"Excell+Good"/"Fair+Poor"  & 0/10 &  ~7/10 ~      &   ~14/4         \\
\hline \hline
\end{tabular}
\end{table}

\begin{figure}[tbh!]
\centerline{
\includegraphics[width=120mm]{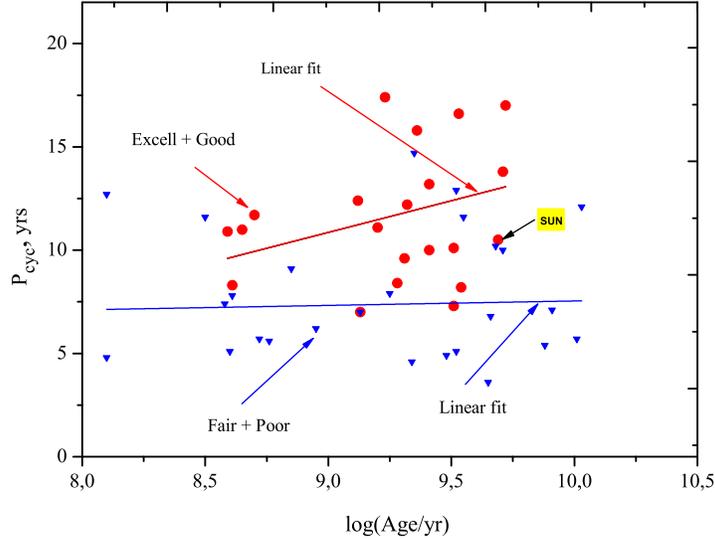}}
 \caption{$P_{cyc}$ versus stellar age.}
{\label{Fi:Fig8}}
\end{figure}

Different tests of the dependency of the cycle period (with
durations in various time scales from seconds in the asteroseismic
analysis to several yrs in Dynamo processes studies) on different
parameters of solar-type stars have been performed, see Morgenthaler
et al. 2006, Garcia et al. 2010, Garcia et al. 2014, Mathur et al.
2012, Metcalfe et al. 2010.

We have analyzed the dependence of the star magnetic cycle duration
on its age. Cycle durations were taken from Baliunas et al. 1995.
Stellar ages were calculated according to Wright et al. 2004 as a
function of chromospheric activity.

In Figure 8, the dependence of $P_{cyc}$ versus age of stars is
shown. The scatter of points around the linear regression line is
very large. For stars with cycles of "Excellent" + "Good" classes
with the increase of the age (or various parameters connected with
the age) the duration of cycles increases by about 20 \% with an
increase of log(Age/yr) from 8.5 to 10. The stars with cycles of
"Fair" + "Poor" classes show no dependence of $P_{cyc}$ on age. So
the problem of determination of $P_{cyc}$ as accurately as possible,
using frequency-time (wavelet) analysis, becomes very actual.

\vskip12pt
\section{Conclusions}
\vskip12pt

\begin{itemize}

\item We believe that the nature of the cyclic activity of solar-type stars is very similar
to the Sun's one: along with main cycles there are quasi-biennial
cycles. Periods of these quasi-biennial cycles evolve during one
main (11-yr cycle) from 2 to 3.5 yrs that complicates their
detection with the periodogram technique. Our conclusion is
supported by the direct observation of cycles with the duration 2-3
yrs for $\tau$ Boo and HD 78366 by Morgenthaler et al. (2011) and
earlier detection of cycles with duration 11.6 and 12.3 yrs by
Baliunas et al. (1995) for the same stars.

\item The quality of the cyclic activity, similar to the solar 11-yr one, is
significantly improved (from "Fair + Poor" to "Excellent + Good") in
G and K-stars as compared to F-stars. The F-stars 11-yr cyclicity
(detected only in every fourth case) is determined with a lower
degree of reliability.

\item The chromospheric activity of HK-project stars is maximal for
stars of the spectral class K, see Table 1. This fact is consistent
with the idea that the chromospheric activity is formed in inner
parts of the stellar convective zone, see Bruevich et al. (2001).
G-stars (and the Sun) are of less chromospheric activity among the
stars studied here, the increased activity of atmospheres of F-stars
is also characterized by enhanced chromospheric activity, slightly
less than that of K-stars. This conclusion is consistent with the
analysis of Isaacson et al. 2010. The Basic chromospheric activity
Level $S_{BL}$  from Isaacson et al. 2010 begins to rise when
$B-V>1$. This increase of $S_{BL}$ is because of a decrease of the
continuum flux for redder stars.

We confirm that in case of chromospheric activity a direct
correlation takes place, in the case of X-ray radiation of stars
versus $(B-V)$ the inverse correlation takes place.

\item The level of chromospheric activity of the Sun is consistent with that of
HK-project stars, which have well-defined cycles of activity ("Excellent + Good") and similar color indexes.
On the other hand, the coronal activity of the Sun is significantly
below that of the coronal activity of G-stars of the HK-project and
other observational Programs.

\item The coronal activity is also more pronounced in stars of the
spectral class F, due to their total increased atmospheric activity
(as compared to stars of spectral classes G and K), and is not
associated with under photospheric convective zones in the practical
absence of chromospheric cycles.

\item Now it's of great interest to find planets of habitable zone
which is the region around a star where a planet with sufficient
atmospheric pressure can maintain liquid water on its surface. We
believe that the close attention should be paid to the unique
characteristics of our Sun: a very low level of variability of the
photospheric radiation simultaneously with a very low level of
coronal radiation.
 Probably, the search for extraterrestrial life should be conducted
simultaneously on the "planets of habitable zone" and on the "stars
comfortable for life", such as the Sun.

\end{itemize}

\end{document}